# Water and molecular chaperones act as weak links of protein folding networks: energy landscape and punctuated equilibrium changes point towards a game theory of proteins

István A. Kovács[1], Máté S. Szalay[1] and Peter Csermely[2]

Department of Medical Chemistry, Semmelweis University,
P.O. Box 260, H-1444 Budapest, Hungary

**Abstract**

Water molecules and molecular chaperones efficiently help the protein folding process. Here we describe their action in the context of the energy and topological networks of proteins. In energy terms water and chaperones were suggested to decrease the activation energy between various local energy minima smoothing the energy landscape, rescuing misfolded proteins from conformational traps and stabilizing their native structure. In kinetic terms water and chaperones may make the punctuated equilibrium of conformational changes less punctuated and help protein relaxation. Finally, water and chaperones may help the convergence of multiple energy landscapes during protein-macromolecule interactions. We also discuss the possibility of the introduction of protein games to narrow the multitude of the energy landscapes when a protein binds to another macromolecule. Both water and chaperones provide a diffuse set of rapidly fluctuating weak links (low affinity and low probability interactions), which allow the generalization of all these statements to a multitude of networks.

**1. Introduction: water and chaperones in protein folding**

The start of protein folding is usually a mixture of two possible scenarios: the initial formation of major elements of the secondary structure and a fast collapse of the unfolded protein, where most of its hydrophobic residues become buried and (more-less) stable intermediates are formed. Later steps of the folding process often include a slow reorganization of the hydrophobic core of the protein. The free energy gain in protein folding enables the presence of local, thermodynamically unstable, "high-energy" protein structures even in the native state, which are stabilized by thermodynamically favorable conformation of all the additional segments of the protein. The "high-energy" local structures can stabilize themselves by forming complexes with another molecule; thus, they often serve as active centers of enzymes or as contact surfaces between various proteins involved *e.g.* in signal transduction [1-5].

Protein folding is accompanied by a massive dehydration [6]. Most of this dehydration occurs during the initial hydrophobic collapse. Though a "wet" folding intermediate would be favorable [7], the expulsion of water molecules seems to be necessary to stabilize the transient structures during and after the initial steps [8-11]. However, water exclusion is not complete: the usual folding intermediate, the molten globule, preserves most of the native internal hydration sites, and has a native-like surface hydration [12]. On the contrary, the intrusion of water molecules and the consequent break in peptide hydrogen bonds is a key step in protein denaturation [10, 13].

---

[1] I.A.K. (steve3281@freemail.hu) and M.S.S. (symat@freemail.hu) are sophomores of the Eötvös Loránd and Technical Universities of Budapest, respectively, and started their research as members of the Hungarian Research Student Association (www.kutdiak.hu), which provides research opportunities for talented high school students since 1996.
[2] To whom all correspondence should be addressed: csermely@puskin.sote.hu (www.weaklinks.sote.hu)

Protein folding often requires assistance. The aggregation of folding intermediates is a great danger, which is prevented by molecular chaperones (passive mode: usually an ATP-independent process). Chaperones recognize and cover hydrophobic surfaces successfully competing with aggregation. However, there is an important difference here. Unlike the aggregating partners, chaperones can release aggregation-prone proteins. This release is usually accompanied by a transfer to another chaperone or occurs after a switch to the other mode of chaperone-mediated folding assistance. The second type of chaperone action (active mode: an ATP-dependent process) is performed by unfolding the incorrectly folded proteins, thus giving them another chance for spontaneous refolding. Passive and active chaperone action is typical to stressful and resting cells, respectively, reflecting the ATP-deficient state under stress. The two types of chaperone-assistance may go in parallel, or may be characteristic of different chaperone-target pairs [14-17].

Setting the exact amount of internal water molecules seems to be a rather important parameter to achieve a "smooth" protein folding [18]. Too much buried water would allow a structural uncertainty, while a complete exclusion of internal water would freeze any further conformational transitions. In agreement with this intricate balance, molecular chaperones were suggested to readjust the internal hydration of proteins. Those chaperones (often called as Anfinsen-cages), which have an internal cavity to sequester partially folded proteins, grab their targets by multiple interactions. An ATP-induced conformational change expands the inner walls of the chaperone-cavity, and stretches, thus partially unfolds the target protein resulting in a preferential mobilization of its internal, hydrophobic core [5, 19-21]. Hydrogen-deuterium studies showed a massive increase in the amount of exchangeable protons during chaperone action. In several models the increased exchange of protons at buried sites of the hydrophobic core of the target protein is accompanied by a rather high level of residual 3D structure. This suggests that chaperones may induce a facilitated entry of additional water molecules to the hydrophobic core of the partially misfolded protein. Thus, certain chaperones may behave as "water-percolators". (Whether real percolation occurs here, thus the chaperone-assisted entry of water molecules leads to a communication of the entire protein structure requires further studies.) As a summary: a class of chaperones may allow the penetration of water molecules to the hydrophobic core of the target protein, and may also allow these water molecules to catalyze conformational transitions necessary to rescue the misfolded target from its folding trap [5].

In the current paper we will describe the above help of water and molecular chaperones in protein folding using the network approach. In part two we will introduce energy and topological networks and show their usefulness to understand the protein folding process. In part three we will summarize earlier suggestions that water and chaperone molecules efficiently smooth the energy landscape of proteins. In part four we will extend this analysis to kinetic events. First we will show that the conformational transitions of proteins may be described as a punctuated equilibrium. Next we will suggest that both water and chaperones make this equilibrium less punctuated. In part five we will return from conformational changes to the analysis of a stable state: the native protein structure and suggest that it is stabilized by water and chaperones. We will raise the idea that water and chaperones may help the convergence of multiple energy landscapes, which arise, when macromolecular complexes are formed. In part six we will extend these thoughts, and will give an initial idea on the formulation of future rules of "protein games" in the currently missing game theoretical approach of these situations. In part seven we will show that both water and chaperones form weak links in protein topological networks, which raises the possibility that all the above statements can be generalized to a multitude of networks at various levels. Finally, in the concluding part we will summarize our ideas and suggest a few points for further studies.

**2. Energy and topological networks of protein folding**

Conformational states of proteins can be efficiently described by energy landscapes. A 3D representation of this landscape is shown in Fig. 1(a). Here the conformational states are reduced to a 2D plane and the energy of each state is shown along the z axis. The landscape approach was suggested by Sewall Wright in 1932 [22] and the powerful concept was applied later to understand protein folding [2, 3, 23, 24]. The energy landscape may be simplified to a network [25, 26]. Here nodes of the network represent local energy minima and links (called energy-links to discriminate them from the chemical bonds of the topological networks) between these energy minima correspond to the transition states between them. The energy landscape of proteins has both a small-world and a scale-free character [25]. The small-worldness of the energy landscape network gives us another explanation, why our proteins fold so efficiently: the node of the network representing the native state is only a few steps (conformational transitions) apart from any other energy minimum of the landscape.

Small-worldness is a typical feature of not only the energy but also the topological networks of proteins [1, 27, 28]. Here, nodes of the network represent segments (atoms, amino acid side chains) of the protein, while links are physical bonds between them as shown in Fig. 1(b). Usually only long-range interactions between amino acid side-chains are taken into consideration when constructing these topological networks. Interestingly, key amino acids, called nucleation centers shown to govern the folding process form highly connected hubs of this small-world network [29]. Moreover, small-world type connectivity increases further during the folding process [30].

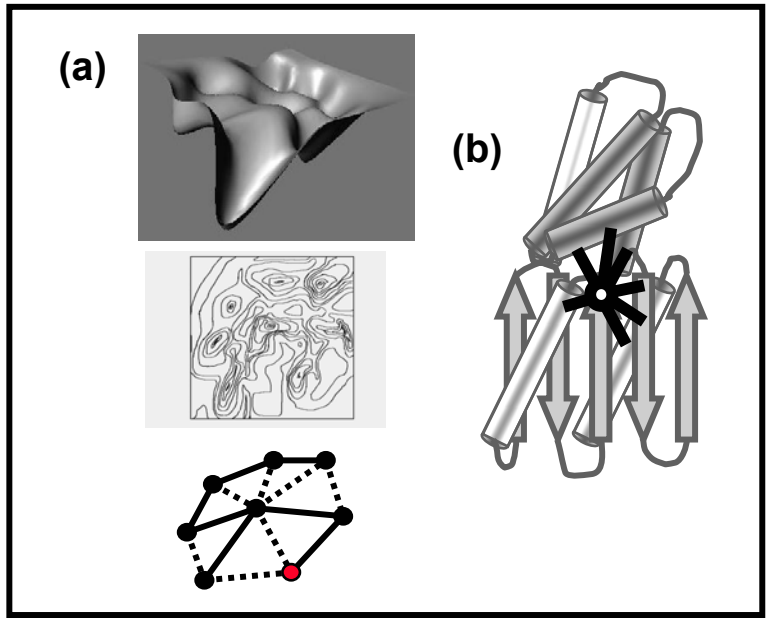

**Figure 1.** (a) Network representation of the energy landscape of protein folding. A hypothetical energy landscape is shown as a 3D image and as a contour plot. On the bottom panel its transformation to a network is described, where the solid and dotted lines represent strong (low activation energy) and weak (high activation energy) energy-links, respectively [23, 24]. (b) Topological network of protein conformation. A node of a hypothetical topological network of the interactions between amino acid side chains is shown after Vendruscolo *et al* [27]. Please note, that these topological links can also be strong and weak depending on the affinity and probability of the bonds, which make them.

## 3. Water and chaperones smooth the energy landscape

Water molecules make fluctuating hydrogen bonds with atoms of the peptide bonds as well as with amino acid side chains [8]. These transient changes induce a fluctuation in the energy level of the actual protein conformation, which open a possibility for a transient decrease in the activation energy between various conformational states. In agreement with these assumptions, a recent paper from Peter Wolynes' lab [31] showed that water efficiently lowers the saddles (activation energies) of the energy landscapes and makes previously forbidden conformational transitions possible. Interestingly, water-fluctuations are decreased in folded states [32], which may indicate a decreased help for protein folding as the multitude of conformational states converges to the native conformation.

Molecular chaperones may act in a similar manner. On one hand, some of the chaperone actions (e.g. that of the active mode Anfinsen-cage chaperones) may capacitate a better access of water molecules to protein segments (e.g. the hydrophobic protein core), where water enters only seldom and to a restricted number of special places called internal cavities [5]. On the other hand, chaperones themselves bind to proteins transiently, and change their binding properties in a highly dynamic manner. In their active mode, the ATP-cycle of molecular chaperones provides a mechanism called "iterative annealing" resulting in an almost continuous change in the actual bonds between the chaperone and the target protein [33]. In the passive (ATP-independent) mode of chaperone action, chaperone binding often uses disorganized segments of the chaperone protein with low affinity interactions [34]. Chaperone-target complexes are transient and here again a significant amount of fluctuations may occur via "stochastic cycling" [35, 36]. In agreement with the above statements for water, chaperone mediated smoothing of the energy landscape has been suggested by Ulrich Hartl and co-workers [37].

## 4. Water and chaperones make the punctuated equilibrium less punctuated

Compared to the energy levels of various protein conformations we know relatively little about the detailed kinetics of conformational transitions of the protein structure. The early work of Ansari et al. [38] showed the existence of "protein-quakes", i.e. the cascading relaxation of myoglobin after the photo-dissociation of carbon monoxide. The protein-quake is an example of the wide-spread self-organized criticality phenomena [39, 40], where an increasing tension is met with a restricted relaxation. As a consequence, sudden avalanches (here: protein quakes) develop, which have a scale-free distribution of both their occurrence and intensity. The scale-free kinetics is related to the scale-free structure of protein surfaces [39] and to the presumed scale-free transport "channels" inside the proteins [42].

The scale-free kinetic character of protein movements means that most conformational changes are restricted and rather small. However, proteins might also take a big jump, which happens only very rarely. If we take these features together with the general notion that self-organized critical phenomena lead to the development of a punctuated equilibrium [43], we may conclude that proteins are in a punctuated equilibrium. In this representation the relative "stasis" corresponds to the fluctuations of the closely related set of conformations around a local (or global) energy minimum, while the "punctuation" means the sudden jump over an activation energy barrier to one of the neighboring stabile conformational states.

If water and chaperones lower the activation energy between two local energy minima as we suggested in the previous part, the transitions between these energy minima should be much less dramatic. Thus the presumed water- and chaperone-induced smoothing of the energy landscape leads to much less punctuation of the punctuated equilibrium.

In agreement with the above assumptions, numerous pieces of experimental evidence indicate that water is an essential and unique source of conformational flexibility of proteins [8, 18]. As an important example of the many, anhydrous enzymes are "frozen", mostly inactive and retain a molecular memory [44, 45]. Fluctuating protein-water interactions allow a better relaxation of the various conformational tensions and thus "dissolve" the large conformational jumps to a set of smaller transitions. This makes the equilibrium less punctuated.

Passive chaperones bind to their unfolded or misfolded target and may help the disorganization of the outer, hydrophobic segments of these proteins [34]. On the contrary, the active chaperone, GroEL increases the internal flexibility of both the RuBisCO [19] and carbonic anhydrase [20] targets. Moreover, Anfinsen-cage chaperones, like GroEL may allow water molecules to enter the hydrophobic core as described above [5]. In agreement with all these findings and assumptions, Hartl and co-workers proposed that chaperones smooth the energy landscape [37]. Thus, chaperones may also make the punctuated equilibrium of protein conformations less punctuated. However, there seems to be a "division of labor" here. While (under chaperone-free, "normal" conditions) water and passive chaperones mobilize mostly the outer segments of proteins, Anfinsen-cage chaperones may help to mobilize the inaccessible internal core [19, 5].

## 5. Water and chaperones may help landscape-convergence

In the above discussions we suggested how water and chaperones help folding or refolding proteins to reach their native conformation by making a smoother transition between energy minima as well as by making the conformational equilibrium less punctuated. Here we show another aspect of the same hypothesis: once the protein has reached the energy minimum, it should stay there to achieve stability. However, time-to-time the protein receives smaller or bigger energy packages from its surroundings or receives direct conformational perturbations. If the energy landscape is rugged (and the conformational equilibrium is punctuated) the native protein may jump to a neighboring, non-native energy state after such a perturbation and its return to the original, native state becomes kinetically inhibited by the large activation energy barrier between the two. Water and chaperones diminish this activation energy barrier, thus cause a kinetic stabilization of the thermodynamically most stable (often but not always: native) structure in the appropriate segment of the energy landscape.

Thinking about the effects of water and chaperones on the energy landscape further, an additional level of complexity emerges. A transiently interacting partner (like water or chaperones) of the protein adds one or many new elements to the topological network described in Fig. 1(b). Extension of this network transiently changes both the conformational space (allowing previously forbidden conformations to occur) and also influences the energy levels. Water and chaperones may not only decrease the activation energy (the saddles) between the energy minima but may also transiently change the x-y plane of the landscape and the absolute value of the energy minima (the z axis). Thus water and chaperones may smooth the energy landscape both by lowering its saddles and by flattening its energy minima due to the averaging of their fluctuating influences. (This may partially explain why dehydration is needed for the completion of the folding process.)

What if stable protein-protein complexes are formed? Non-transient, high-probability and usually rather high affinity, strong interactions occur. These interactions may dramatically change the original energy landscapes of the participating proteins. At the binding event a rather complex set of "protein-quakes" is expected to happen as the two proteins try to relax the complex perturbations. Each element of the relaxation events of any interacting protein changes the energy landscape of both partners. The parallel search for conformational optima in the mutually and continuously changing energy landscapes of the interacting partners would be cognitively and computationally intractable for the experimenters, and for the participating proteins themselves would probably require an astronomical time without the help of water and (quite often) chaperones. In addition to their presumed help in the positioning and steering of the two binding partners towards each other [14, 31], water and chaperones may also help the solution of the "conformational conflict" of their initial contacts. In agreement with these ideas water has been suggested to help avoiding the conformational frustrations by Papoian and Wolynes [46].

As a summary of our hypothesis: water and chaperones may not only bring both the proteins and their energy landscapes into motion but may also help the convergence of the various energy landscapes after a protein-protein, protein-membrane, protein-RNA or protein-DNA interaction. Without water and chaperones the assembly of macromolecular complexes would be hopelessly inefficient and self-organization would not easily occur. This approach gives an additional element to the necessity (paramount influence) of water for the development of the current form of life on Earth. Moreover, the same logic extends the previous ideas [47, 48] on the importance of molecular chaperones in the early stages of evolution.

**6. Initial thoughts on protein games**

The above scenario of protein-macromolecule interactions raises yet an additional approach. The cognitively (and numerically) intractable complexity of the situations, when the elements themselves continuously influence each other's landscapes (like the evolutionary landscape, innovation landscape, etc.) is usually approached by the game theory. The application of game theory in the first approximation requires two actors who consciously recognize the other's actions and modify the set and preference of their potential responses (the points of the x-y plane as well as the positioning of these points on the z axis of their "energy landscape" as shown on figure 1). In this strict sense of games the application of the game theory to macromolecular interactions becomes impossible. However, if we take into account that the complexity of the topological networks of proteins already results in a mutual distortion of the original energy landscape of both interacting partners, which – in turn – changes their interactions and therefore a sequence of complex events occur as described above, the necessity and possibility of the introduction of "protein game rules" becomes conceivable. If our proteins were complex enough to be conscious (as they are obviously not), the situation would be even more complex, since only a perception but no physical interaction (geometrical proximity) would be required to change their landscapes upon the observation of other players' acts or preferences in the game. We are lucky: this will never happen. Proteins will not have morning news announcing a change in the attitude of the cell they happen to inhabit.

However, protein complexes are still sophisticated enough to require the drastic but highly rational simplifications of the game theory in their modeling.

In Table 1 we suggest a set of criteria to classify protein-protein interactions. A few elements of these aspects have been already characterized, thus steering was already discriminated to electrostatic interaction- [49] or desolvation-mediated [50] mechanisms, and initial binding may often involve specific anchor residues [51].

Table 1
How does the key get to the lock and induced fit get induced? Preliminary aspects of the classification of future protein games

| Aspect | Possible scenarios |
| --- | --- |
| 1. Where do the two proteins bind to each other? | |
| geometry of the binding surface | "flat" |
| | convex/concave (degree of binding area complexity, fractal dimension, total surface area) |
| binding surface localization | few amino acid side chains |
| | complex domain |
| | multiple domains |
| 2. How do the two proteins bind to each other? | |
| steering (approaching each other) | no steering at all: bouncing or binding |
| | steering by weak protein-protein interactions |
| | steering by water-mediated contacts |
| folding | both fold |
| | both unfold |
| | one of them folds the other unfolds |
| | one of them folds the other is unchanged |
| | one of them unfolds the other is unchanged |
| | none of them is changed |
| cooperation | both proteins are "rigid": no cooperation during binding |
| | proteins are "flexible": cooperation |
| binding trajectory | both energy landscapes are rough |
| | one energy landscape is rough the other is smooth |
| | both energy landscapes are smooth |
| water | help of water is essential |
| | help of water is important |
| | help of water is negligible |
| other actor(s) | only amino acid side chains participate |
| | non-pertinacious, covalently bond elements (oligosaccharides, lipids, etc.) are on the binding surface |
| | co-factors participate (e.g. binding needs chaperones) |
| 3. What holds the two proteins together? | |
| type of bonds | a few unique bonds (e.g. ion-pairs) |
| | disperse bonds (H-bridges, van der Waals forces) |
| | both |
| bond strength | Strong |
| | weak |
| | both |

The introduction of protein games becomes justified, if the development of protein-protein interactions can be classified to well-discriminated basic scenarios, and these scenarios are highly populated in the multitude of possible energy landscape-pairs. Domain-swapping (when two secondary structures or whole domains are exchanged between the binding partners through a simultaneous unfolding/refolding event) [52] and fly-casting (when one of the partners folds as the interaction develops) [53] are all possible scenarios of future protein games. We have to note that in the cellular context binding of two proteins is modulated by a whole range of interactions with other elements of larger protein complexes, etc., which make the game-rules even more complex. As an additional note, we are also aware of the fact that the above thoughts on protein games can be easily extended to any protein-macromolecule (DNA, RNA, membrane, etc.) interactions.

## 7. Water and chaperones as weak links

Both water molecules and chaperones form low affinity, low probability weak links with the segments of proteins they interact with [32, 36]. Therefore, we can formulate the above assumptions in the following way: weak links may generally lower the saddles of energy landscapes and by allowing a faster relaxation, make the conformational equilibrium less punctuated. Weak links develop in the final (latching) steps of anchor-mediated protein binding [51], and during the constant reorganization of the fly-casting binding mechanism [53]. Symmetric homodimers are often formed by two non-identical monomers, where one of the monomers is more folded than the other, and serves as a template for the folding of the other [54]. Here again, diversity introduces a set of stabilizing weak links. An optimal amount of weak links may also induce the convergence of the energy-type landscapes in case of network-network interactions helping the simplification of most if not all scenarios usually treated by the game theory [26, 55].

## 8. Summary and perspectives

The most important hypotheses of the current paper can be summarized as follows:
- water and chaperones may allow a faster relaxation of tensions in protein conformations thus may make the occurrence of protein-quakes (avalanches of self organized protein criticality) less likely; in other words: proteins are in a punctuated equilibrium of conformational states, water and chaperones may make this conformational equilibrium less punctuated; as a consequence: water and chaperones may stabilize the local or global minimum energy (often: native) state;
- water and molecular chaperones may smooth the energy landscape of proteins and may develop easier transitions towards the completion of the folding process [31, 37]; as a next step: water and chaperones may allow the convergence of multiple energy landscapes in case of protein-macromolecule interactions [46]; the introduction of a protein game theory may be useful to point out and understand the most frequent binding mechanisms from the multitude of possible scenarios;
- water and chaperones behave like weak topological links in the protein structure network and establish weak energy-links in the energy network of the landscape model. This may allow the generalization of all the above statements to a multitude of social, ecological and other networks having weak links [55].

The role of water and chaperones in the conformational transitions of proteins can be experimentally tested by measuring the kinetics of protein relaxation after a perturbation in the presence and absence of chaperones or various availability of water. If our assumptions are correct, "smoother" transitions and much less "protein-quakes" should be observed, if water is freely available and chaperones are present. The justification of protein games can be tested by systematic analysis of protein-complex databases [46] and by the assessment, if binding scenarios populate rather separated islands in the energy landscape continuum. The generalization of weak link-induced effects raises a multitude of exciting questions, which go way beyond the scope of the current paper and will be described in the upcoming book of one of the authors [55]. The interface (fringe area) between the network approach and protein folding studies is a very rich field, which may help us to develop a novel understanding of both.

**Acknowledgments**

The authors would like to thank members of the LINK-group (www.weaklinks.sote.hu) for helpful discussions and to Peter Tompa (Institute of Enzymology, Hungarian Academy of Sciences, Budapest, Hungary) for critical reading of the manuscript. Work in the authors' laboratory was supported by research grants from the EU (FP6-506850, FP6-016003), Hungarian Science Foundation (OTKA-T37357 and F47281), Hungarian Ministry of Social Welfare (ETT-32/03) and by the Hungarian National Research Initiative (NKFP-1A/056/2004 and KKK-0015/3.0).